\documentclass[final,5p,times,twocolumn]{elsarticle}

\usepackage[
pdftitle={Physical Complexity of Variable Length Symbolic Sequences},
pdfauthor={Gerard Briscoe},
pdfsubject={Physical Complexity},
pdfkeywords={Evolution, complexity, entropy, clustering, population},
colorlinks=true, linkcolor=black, citecolor=black, filecolor=black, urlcolor=black, hyperfootnotes=false]{hyperref}

\usepackage{graphicx, acronym, ifthen, substr, color}

\usepackage{amsthm, amsmath}

\newboolean{final}\setboolean{final}{true}

\newboolean{mactex}\setboolean{mactex}{true}

\newcommand{\added}[1]{#1}
\newcommand{\changed}[1]{#1}

\newboolean{arxiv}\setboolean{arxiv}{false}

\ifthenelse{\boolean{final}}{}{\usepackage{showlabels}\usepackage{citeext}}

\newcommand{\tfigure}[9]
    {
    \IfSubStringInString{!}{#7}{\begin{figure}[#7]}{\begin{figure}[!t]}
    \IfSubStringInString{mm}{#8}{\vspace{#8}}{}
    \centering

    \IfSubStringInString{pdf}{#3}
        {
        \execute{cd images; ln -s #2.pdf .#2.gdf}
        \includegraphics[#1]{images/#2}
        }
        {
        \execute{cd images; ./pdfcrop.sh #2}
        \includegraphics[#1]{images/#2-crop.pdf}
        }
    \vspace{#6}
    \caption[#4]
        {
        \label{#2}
        #4: #5
        }
    \IfSubStringInString{mm}{#9}{\vspace{#9}}{}
    \end{figure}
    }

\definecolor{tred}{RGB}{255,0,0}
\def\tred#1{#1}

\newcommand{\squareG}[3]
    {
    \squaresub{#1}{#2}{#3}{-1pt}
    }

\newcommand{\squaresub}[4]
    {
    \immediate\write18{cd images; ./pdfcrop.sh square#2}
    \ifthenelse{\boolean{final}}
        {\hspace{#1}\raisebox{#4}{$\includegraphics[scale=0.75,clip=true, trim=0mm 0.25mm 0.25mm 0mm]{images/square#2-crop.pdf}$}\hspace{#3}}
        {\href{file://localhost/Users/g/Desktop/PhDthesis/images/square#2.graffle}{\hspace{#1}\raisebox{#4}{$\includegraphics[scale=0.75,clip=true, trim=0mm 0.25mm 0.25mm 0mm]{images/square#2-crop.pdf}$}\hspace{#3}}}
    }

\newcommand{\execute}[1]{\immediate\write18{#1}}

\definecolor{tred}{RGB}{255,0,0}

\newcommand{\setCap}[2]{#1\immediate\write18{./mkcaption.sh #2}}
\newcommand{\getCap}[1]{\acl*{#1}}

\acrodef{PCG}{Projected Conjugate Gradient} 
\acrodef{QP}{quadratic programming}
\acrodef{RBF}{Radial-Basis Function}
\acrodef{ABM}{Agent-Based Modelling}
\acrodef{AI}{Artificial Intelligence}
\acrodef{DAI}{Distributed Artificial Intelligence}
\acrodef{API}{Application Programming Interface}
\acrodef{ARF}{p14ARF human tumor-suppressor gene}
\acrodef{B2B}{business-to-business}
\acrodef{BDP}{Biological Design Pattern}
\acrodef{BGS}{Best Guess Solution}
\acrodef{BIC}{Biologically-Inspired Computing}
\acrodef{BML}{Business Modelling Language}
\acrodef{BPEL}{Business Process Execution Language}
\acrodef{BPMN}{Business Process Modelling Notation}
\acrodef{CAS}{Complex Adaptive Systems}
\acrodef{COBOL}{COmmon Business-Oriented Language}
\acrodef{DBE}{Digital Business Ecosystem}
\acrodef{DE}{Digital Ecosystem}
\acrodef{DEC}{distributed evolutionary computing}
\acrodef{DGA}{Distributed genetic algorithms}
\acrodef{DIS}{Distributed Intelligence System}
\acrodef{DNA}{Deoxyribose Nucleic Acid}
\acrodef{DOP}{DBE Open Protocol}
\acrodef{DSS}{Distributed Storage System}
\acrodef{EAP}{Evolving Agent Population}
\acrodef{ebXML}{e-business eXtensible Markup Language}
\acrodef{EC}{Evolutionary Computing}
\acrodef{ECJ}{Evolutionary Computing in Java}
\acrodef{EE}{Evolutionary Environment}
\acrodef{EFL}{Evolutionary Framework for Language}
\acrodef{FLE}{Framework for Language Ecosystems}
\acrodef{EOA}{Ecosystem-Oriented Architecture}
\acrodef{ESS}{evolutionary stable strategy}
\acrodef{EvE}{Evolutionary Environment}
\acrodef{ExE}{Execution Environment}
\acrodef{FCB}{Framework for Computational Biomimicry}
\acrodef{FFF}{Fitness Function Framework}
\acrodef{FL}{Fitness Landscape}
\acrodef{HWU}{Heriot-Watt University}
\acrodef{ICL}{Imperial College London}
\acrodef{ICT}{Information and Communications Technology}
\acrodef{INTEL}{Intel Ireland}
\acrodef{IPA}{International Phonetic Alphabet}
\acrodef{ISUFI}{Istituto Superiore Universitario di Formazione Interdisciplinare}
\acrodef{JDJ}{Java Developer's Journal}
\acrodef{KC}{Kolmogorov-Chaitin}
\acrodef{LAN}{local area network}
\acrodef{LSE}{London School of Economics and Political Science}
\acrodef{MAS}{Multi-Agent System}
\acrodef{MDL}{Minimum Description Length}
\acrodef{MDM2}{murine double minute 2}
\acrodef{MFT}{Mean Field Theory}
\acrodef{MoAS}{Mobile Agent System}
\acrodef{MOF}{Meta Object Facility}
\acrodef{MUH}{migration and usage history}
\acrodef{NIC}{Nature Inspired Computing}
\acrodef{NN}{Neural Network}
\acrodef{NoE}{Network of Excellence}
\acrodef{OMG}{Open Mac Grid}
\acrodef{OPAALS}{Open Philosophies for Associative Autopoietic Digital Ecosystems}
\acrodef{P2P}{peer-to-peer}
\acrodef{P53}{protein 53}
\acrodef{PDA}{Personal Digital Assistant}
\acrodef{QoS}{quality of service}
\acrodef{REST}{REpresentational State Transfer}
\acrodef{RNA}{Deoxyribose Nucleic Acid}
\acrodef{SAE}{Software Agent Ecosystem}
\acrodef{SBML}{Systems Biology Modelling Language}
\acrodef{SBVR}{Semantics of Business Vocabulary and Business Rules}
\acrodef{SDL}{Service Description Language}
\acrodef{SF}{Service Factory}
\acrodef{SIM}{Social Interaction Mechanism}
\acrodef{SM}{Service Manifest}
\acrodef{SME}{Small and Medium sized Enterprise}
\acrodef{SML}{Service Modelling Language}
\acrodef{SMO}{Sequential Minimal Optimisation}
\acrodef{SOA}{Service-Oriented Architecture}
\acrodef{SOAP}{Simple Object Access Protocol}
\acrodef{SOC}{Self-Organised Criticality}
\acrodef{SOLUTA}{SOLUTA.NET}
\acrodef{SOM}{Self-Organising Map}
\acrodef{SSL}{Semantic Service Language}
\acrodef{STU}{Salzburg Technical University}
\acrodef{SUN}{Sun Microsystems}
\acrodef{SVM}{Support Vector Machine}
\acrodef{TM}{Turing Machine}
\acrodef{UBHAM}{University of Birmingham}
\acrodef{UDDI}{Universal Description Discovery and Integration}
\acrodef{UML}{Unified Modelling Language}
\acrodef{URI}{Uniform Resource Identifier}
\acrodef{UTM}{Universal Turing Machine}
\acrodef{VLP}{variable length population}
\acrodef{VLS}{variable length sequences}
\acrodef{vls}{variable length sequence}
\acrodef{WP}{Work-Package}
\acrodef{WSDL}{Web Services Definition Language}
\acrodef{XMI}{XML Metadata Interchange}
\acrodef{XML}{eXtensible Markup Language}
\acrodef{MD5}{Message-Digest algorithm 5}
\acrodef{GA}{genetic algorithm}
\acrodef{GP}{genetic programming}
\acrodef{MASON}{Multi-Agent Simulator Of Neighbourhoods}
\acrodef{Repast}{Recursive Porous Agent Simulation Toolkit}
\acrodef{JCLEC}{Java Computing Library for Evolutionary Computing}
\acrodef{OWL-S}{Web Ontology Language - Service}
\acrodef{EGT}{Evolutionary Game Theory}
\acrodef{RBF}{Radial Basis Functions}
\acrodef{SWS}{Semantic Web Services}
\acrodef{HDD}{Hard Disk Drive}
\acrodef{SSD}{Solid-State Drive}

\acrodef{genCap2}{equalling the maximum length would be incorrect}
\acrodef{orgCPcap}{are consistent with the intuitive understanding one would have for the complexity of the sample populations}
\acrodef{FLsingle2Cap}{The sequences of an evolving population will evolve, clustering around the optimal genome}
\acrodef{FLmul2Cap}{Efficiency $E$ will tend to a maximum below one, because the population of sequences consists of more than one cluster, with each having an Efficiency tending to a maximum of one.}
\acrodef{FLmulCap}{The simplest scenario of clusters is {pure clusters}}
\acrodef{popClusShowCap}{The clusters of the population have Efficiency values tending to a maximum of one, compared to the Efficiency of the population as a whole, which is tending to a maximum significantly below one.}
\acrodef{largeVisCap}{The visualisation shows that our Efficiency $E$ accurately measures the complexity of the two populations.}
\acrodef{graph32cap}{The Efficiency tends to a maximum of one, indicating that the population consists of one cluster}
\acrodef{graph4cap}{around the included best fit curve, quite significantly at the start, and then decreasing as the generations progressed.}
\acrodef{visClustersCap}{sequences were grouped to show the two clusters}
\acrodef{visClusters2Cap}{expected from (\ref{defineCluster}) each cluster had a much higher Physical Complexity and Efficiency compared to the population as a whole. However, the Efficiency $E_{c}$}
\acrodef{visClusters3Cap}{calculated the}

\newcommand{\white}[1]{\color{white}#1\normalcolor}

\newcommand{\be}{\begin{equation}}
\newcommand{\eeq}[1]{\label{#1}\end{equation}}

\begin{document}

\begin{frontmatter}

\title{Physical Complexity of Variable Length Symbolic Sequences}

\author[label1]{Gerard Briscoe}

\author[label2]{Philippe De Wilde}

\address[label1]{Systems Research Group, Computer Laboratory, University of Cambridge, United Kingdom, gerard.briscoe@cl.cam.ac.uk}

\address[label2]{Intelligent Systems Lab, Department of Computer Science, Heriot Watt University, United Kingdom, p\tred{.de\_wilde}@hw.ac.uk}

\begin{abstract}

A measure called Physical Complexity is established and calculated for a population of sequences, based on statistical physics, automata theory, and information theory. It is a measure of the quantity of information in an organism's genome. It is based on Shannon's entropy\tred{,} measur\tred{ing} the information in a population evolved in its environment\tred{,} by using entropy to estimate the randomness in the genome. It is calculated from the difference between the maximal entropy of the population and the actual entropy of the population when in its environment, estimated by counting the number of \tred{fixed} \emph{loci} in the sequences of a population. Up to now, Physical Complexity has only been formulated for populations of sequences with the same length. Here, we investigate an extension to support \emph{variable} length populations. We then build upon this to construct a measure for the efficiency of information storage, which we later use in understanding clustering within populations. Finally, we investigate our extended Physical Complexity through simulations, show\tred{ing it to b}e consistent with the original\tred{.}

\end{abstract}

\begin{keyword}
\changed{complexity \sep entropy \sep clustering \sep evolution \sep population}
\end{keyword}

\end{frontmatter}

\section{Introduction}
Physical Complexity was born \cite{adami1998ial} from the need to determine the proportion of information in sequences of DNA, because it has long been established \cite{thomasjr1971goc} that the information contained is not directly proportional to the length, known as the C-value enigma/paradox \cite{gregory2001cco}. However, because Physical Complexity analyses an ensemble of DNA sequences, the consistency between the different solutions shows the information, and the differences the redundancy \cite{adami2003}. Entropy, a measure of disorder \cite{vonbertalanffy1973gst}, is used to determine the redundancy from the information in the ensemble (populations). Physical Complexity therefore provides a context-relative definition for the complexity of a population without needing to define the context (environment) explicitly \cite{adami2000}.
\added{It is widely recognized that complexity is best measured using thermodynamic depth. Lloyd and Pagels make this case in \cite{lloyd1988complexity}. The introduction of thermodynamic depth allows them to prove ``the average complexity of a state must be proportional to the Shannon entropy of the set of trajectories that experiment determines can lead to that state''. They further relate thermodynamic depth to computational complexity, and derive a lower bound from the mutual information. We follow a slightly different route, following \cite{adami2000,adami2000-14}, but the Shannon entropy also features in our complexity definition (\ref{avgCompPart1}).}

Physical Complexity is currently formulated for a population of \emph{same length} symbolic sequences \cite{adami2000}. So, we decided to extend Physical Complexity to include populations of \emph{variable length} symbolic sequences, and therefore allow for its improved and wider applicability \cite{bionetics, thesis, phycom}.\\\\\\

\section{Physical Complexity}
\label{measureSelfOrg}

Physical Complexity was derived \cite{adami2000} from the notion of \emph{conditional complexity} defined by Kolmogorov, which is different from traditional Kolmogorov complexity and states that the determination of complexity of a sequence is conditional on the environment in which the sequence is interpreted \cite{li1997ikc}. In contrast, traditional \ac{KC} complexity is only conditional on the implicit rules of mathematics necessary to interpret a programme on the tape of a \ac{TM}, and nothing else \cite{li1997ikc}. So, if we consider a \ac{TM} that takes a tape $e$ as input (which represents its physical environment), including the particular rules of mathematics of this \emph{world}; without such a tape, this \ac{TM} is incapable of computing anything, except for writing to the output what it reads in the input. Thus, without tape $e$ all sequences $s$ have maximal \ac{KC}-complexity, because there is nothing by which to determine regularity \cite{adami2000}. However, \emph{conditional complexity} can be stated as the length of the smallest programme that computes sequence $s$ from an environment $e$,
\begin{equation}
K(s|e) = \min \left\{ {|p|:s = C_T(p,e)} \right\},
\label{komCom}
\end{equation}
where $C_T(p,e)$ denotes the result of running programme $p$ on \acl{TM} $T$ with the input sequence $e$ \cite{adami2000}. This is not yet Physical Complexity, but rather, it is the smallest programme that computes the sequence $s$ from an environment $e$, in the limit of sequences of infinite length, containing only the bits that are entirely unrelated to $e$, since, if they were not, they could be obtained from $e$ with a programme of a size tending to zero \cite{adami2000}. The Physical Complexity $K(s:e)$ can now be defined as the number of bits that are meaningful in sequence $s$ (that can be obtained from $e$ with a programme of vanishing size), and is given by the \emph{mutual complexity} \cite{kolmogorov},
\begin{equation}
K(s:e) = K(s|\emptyset ) - K(s|e),
\label{komNot}
\end{equation}
where $K(s|\emptyset)$ is the unconditional complexity with an empty input tape, $e \equiv \emptyset$ \cite{adami2000}. This is different from the Kolmogorov complexity, because in Kolmogorov's construction the rules of mathematics were given to the \ac{TM} \cite{li1997ikc}. As argued above, every sequence $s$ is random if no environment $e$ is specified, as non-randomness can only exist for a specific world or environment. Thus, $K(s|\emptyset )$ is always maximal,
\begin{equation}
K(s|\emptyset ) = |s|,
\label{komEmpty}
\end{equation}
and is given by the length of $s$ \cite{adami2000}. So (\ref{komNot}) represents the length of the sequence $s$, minus those bits that cannot be obtained from $e$. So, conversely (\ref{komNot}) represents the number of bits that can be obtained in a sequence $s$, by a computation with vanishing programme size, from $e$. Thus, $K(s: e)$ represents the Physical Complexity of $s$ \cite{adami2000}. The determination of the Physical Complexity, $K(s: e)$, of a sequence $s$ with a description of the environment $e$ is not practical. Meaning that it cannot generally be determined by inspection, because its impossible to determine which, and how many, of the bits of sequence $s$ correspond to information about the environment $e$. The reason is that we are generally unaware of the coding used to code information about $e$ in $s$, and therefore coding and non-coding bits look entirely alike \cite{adami2000}. However, it is possible to distinguish coding from non-coding bits if we are given multiple copies of sequences that have adapted to the environment, or more generally, if a statistical ensemble (population) of sequences is available to us. Then, coding bits are revealed by non-uniform probability distributions across the population (\emph{conserved sites}), whereas random bits have uniform distributions (\emph{volatile sites}) \cite{adami2000}. The determination of complexity then becomes an exercise in information theory, because the average complexity $\langle K \rangle$, in the limit of infinitely long strings, tends to the entropy of the ensemble of strings $S$\footnote{This holds for near-optimal codings. For strings $s$ that do not code perfectly we have $\langle K \rangle \ge H$ \cite{zurek1990aic}.} \cite{adami2000-14},
\begin{equation}
 \left\langle {K(s)} \right\rangle _S = \sum\limits_{s \in S} {p(s)K(s)} \approx H(S),
\label{avgCompPart1}
\end{equation}
where $H$ is defined from Shannon's (information) entropy \cite{mackay}, and is given by
\begin{equation}
H(S)=log_{n}(S),
\end{equation}
where $n$ is the number of symbols available for encoding. If each symbol is equally probable, we can rewrite the above function as
\begin{eqnarray}
H(S)&=&-log_{n}(1/S) \nonumber \\
&=&-log_{n}(p),
\end{eqnarray}
where $p$ is the probability of occurrence of any one of the symbols. For a source that outputs an infinite sequence of bits, to communicate a finite set of symbols $S$, Shannon generalised the above function to express an average symbol length \cite{mackay}. This derivation is easier to see for a large, but finite, number of symbols $N$,
\begin{eqnarray}
 H(S) &=& \frac{{\sum\limits_{i = 1}^S {N_i \left[ { - \log_{N} (1/S_i )} \right]} }}{{\sum\limits_{i = 1}^S {N_i } }} = \frac{{\sum\limits_{i = 1}^S {N_i \left[ { - \log_{N} (1/S_i )} \right]} }}{N} \nonumber \\
\white{.} & \white{.} \nonumber \\
&=& - \sum\limits_{i = 1}^S {\frac{{N_i }}{N}\left[ {\log_{N} (1/S_i )} \right]} = - \sum\limits_{i = 1}^S {p_i \log_{N} (p_i )},
\label{Hderivation}
\end{eqnarray}
where $N_i$ is the number of occurrences of the symbol $S_i$. So, given (\ref{avgCompPart1}) and (\ref{Hderivation}), the \emph{average complexity} of the sequences $s$ of a population $S$, $\left\langle {K(s)} \right\rangle _S$, tends to the entropy of the sequences $s$ in the ensemble $S$ \cite{adami2000},
\begin{equation}
 \left\langle {K(s)} \right\rangle _S = -\sum\limits_{s \in S} {p(s)\log p(s)}.
\label{avgComp}
\end{equation}

(\ref{avgComp}) remains consistent with (\ref{komEmpty}) as the determination of $K(s|\emptyset )$, sequence $s$ without an environment $e$, must equal the sequence's length $|s|$, because Shannon's formula for entropy is an average logarithmic measure of the symbol sets \cite{mackay}, and so the maximum entropy of a population is equivalent to the length of the sequences in the population, $H_{max}(S) = |s|$. Indeed, if nothing is known about the environment to which a sequence $s$ pertains, then according to the \emph{principle of indifference}\footnote{The \emph{principle of indifference} states that if there are $n>1$ mutually exclusive and collectively exhaustive possibilities, which are indistinguishable except for their names then each possibility should be assigned an equal probability $\frac{1}{n}$ \cite{jaynes2003ptl}.}, the probability distribution $p(s)$ must be uniformly random. However, if an environment $e$ is given we have some information about the system, and the probability distribution will be non-uniform. Indeed, it can be shown that for every probability distribution $p(s|e)$, to find sequence $s$ given environment $e$, we have
\begin{equation}
H(S|e) \le H(S|\emptyset ) = |s|,
\label{shannonEntropy}
\end{equation}
because of the concavity of Shannon entropy \cite{adami2000}. So, the difference between the maximal entropy $H(S|\emptyset ) = |s|$ and $H(S|e)$, according to the construction outlined above, represents the average number of bits in sequence $s$ taken from the population $S$ that can be obtained by zero-length universal programmes from the environment $e$. Therefore, the average mutual complexity of sequences $s$ in a population $S$, given an environment $e$, is
\begin{eqnarray}
\left\langle {K(s:e)} \right\rangle _S &=& \sum\limits_{s \in S} {p(s)K(s:e)} \nonumber \\
&\approx& H(S|\emptyset ) - H(S|e) \nonumber \\
&\equiv& I(S|e),
\label{avgInfo}
\end{eqnarray}
where $I(S|e)$ is the information about the environment $e$ stored in the population $S$, which we identify with the Physical Complexity \cite{adami2000}. To estimate $I(S|e)$ it is necessary to estimate the entropy $H(S|e)$ using a representative population of sequences $S$ for a given environment $e$, by summing, over the sequences $s$ of the population $S$, the probability $p(s|e)$ multiplied by the logarithm of the probability $p(s|e)$,
\begin{equation}
H(S|e) = - \sum\limits_{s \in S} {p(s|e)\log p(s|e)}.
\label{popEntropy}
\end{equation}
The entropy $H(S|e)$ can be estimated by summing the per-site $H(i)$ entropies of the sequence,
\begin{equation}
H(S|e) \approx \sum\limits_{i = 1}^{|s|} {H(i)},
\label{estEntropySumSite}
\end{equation}
where $i$ is a site in the sequence $s$ \cite{adami2000}. Random sites are identified by a nearly uniform probability distribution, and contribute positively to the entropy, whereas non-random sites (which have strongly peaked distributions) contribute very little \cite{adami2000}. So, the Physical Complexity, the average mutual complexity of sequences $s$ in a population $S$ for an environment $e$, $\left\langle {K(s:e)} \right\rangle _S$, abbreviated as $C$, is the maximal entropy $H(S|\emptyset )$ minus the sum of the per-site entropies,
\begin{equation}
C = H(S|\emptyset ) - \sum\limits_{i = 1}^{|s|} {H(i)}.
\label{avgCom}
\end{equation}
If the sequences $s$ are constructed from an alphabet, a set $D$, then the per site entropy $H(i)$ for the sequences is
\begin{equation}
H(i) = - \sum\limits_{d \in D} {p_d (i)\log _{|D|} p_d (i)}, \\
\label{persite}
\end{equation}
where $i$ is a site in the sequences ranging between one and the length of the sequences $\ell$, $D$ is the alphabet of characters found in the sequences, and $p_d(i)$ is the probability that site $i$ (in the sequences) takes on character $d$ from the alphabet $D$, with the sum of the $p_d(i)$ probabilities for each site $i$ equalling one, $\sum\limits_{d \in D} {p_d (i) = 1}$ \cite{adami2000}. Taking the log to the base $|D|$ conveniently normalises $H(i)$ to range between zero and one,
\begin{equation}
0 \le H(i) \le 1.
\label{persiteMinMax}
\end{equation}
If the site $i$ is identical across the population it will have no entropy,
\begin{equation}
H_{\min } (i) = 0.
\label{persiteMin}
\end{equation}
If the content of site $i$ is uniformly random, i.e. the $p_d (i)$ probabilities all equal to $\frac{1}{{|D|}}$, it will have maximum entropy,
\begin{equation}
H_{\max } (i) = 1.
\label{persiteMax}
\end{equation}
When the entropy of $H(i)$ is at its minimum of zero, then the site $i$ holds information, as every sample shows the same character of the alphabet. When the entropy of $H(i)$ is at its maximum of one, the character found in the site $i$ is uniformly random and therefore holds no information. So, the amount of information is the maximal entropy of the site (\ref{persiteMax}) minus the actual per-site entropy (\ref{persite}) \cite{adami2000},
\begin{eqnarray}
 I(i) &=& H_{\max } (i) - H(i) \nonumber \\
 &=& 1 - H(i) .
\label{persiteInfo}
\end{eqnarray}

For clarity the length of the sequences $|s|$ will be abbreviated to $\ell$ \cite{adami2000},
\begin{equation}
|s| \equiv \ell .
\label{clarity}
\end{equation}
So, the complexity of a population $S$, of sequences $s$, is the maximal entropy of the population (equivalent to the length of the sequences) $\ell$, minus the sum, over the length $\ell$, of the per-site entropies $H(i)$,
\begin{equation}
C = \ell - \sum\limits_{i = 1}^\ell {H(i)},
\label{complexity}
\end{equation}
given (\ref{avgCom}), (\ref{shannonEntropy}) and (\ref{clarity}) \cite{adami2000}. The equivalence of the maximum complexity to the length matches the intuitive understanding that if a population of sequences of length $\ell$ has no redundancy, then their complexity is their length $\ell$.

If $G$ represents the set of all possible genotypes constructed from an alphabet $D$ that are of length $\ell$, then the size (cardinality) of $|G|$ is equal to the size of the alphabet $|D|$ raised to the length $\ell$,
\begin{equation}
|G| = |D|^\ell.
\label{recPopSize}
\end{equation}
For the complexity measure to be accurate, a    sample size of $|D|^\ell$ is suggested to minimise the error \cite{adami2000, basharin}, but such a large quantity can be computationally infeasible. The definition's creator, for practical applications, chooses a population size of $|D|\ell $, which is sufficient to show any trends present. So, for a population of sequences $S$ we choose, with the definition's creator, a computationally feasible population size of $|D|$ times $\ell $,
\begin{equation}
|S|\ \ge \ |D|\ell.
\label{popSize}
\end{equation}
The size of the alphabet, $|D|$, depends on the domain to which Physical Complexity is applied. For example, the alphabet of RNA is its four nucleotides, $D = \{A, C, G, U\}$, and therefore $|D|=4$ \cite{adami2000}.

\added{Our complexity measure is inexorably dependent on Shannon entropy, because of the results in \cite{adami2000-14} that lead to equation (\ref{avgCompPart1}). The symbolic sequences that we analyse are encoding computer programmes that are evolved using evolutionary dynamics \cite{bionetics, thesis, phycom, de07oz, dbebkpub, acmMedes, epi}. We have no a-priori knowledge about the statistics of the sequences. If we were analyzing real DNA or financial time series, we would have knowledge of nonstationarity on different length scales, L\'evy distributions, or other features of non-well behaved systems. If that was the case, it would be appropriate to consider Jensen-Shannon divergence in its regular \cite{grosse2002analysis} or non-logarithmic version \cite{lamberti2003non}. The latter uses Tsallis entropy \cite{tsallis2003nonextensive}. This has had impressive applications in financial time series \cite{duartequeiros1,duartequeiros2,martin1999fisher}. Gell-Mann and Lloyd \cite{gell2003effective} also point out applications in systems on the edge of chaos. If we could prove that our strings encoding computer programmes (multi-agent systems) had similar abnormal statistics, we would use measures based on non-logarithmic entropy. However, in the absence of such information, we will follow Adami \cite{adami2000} in his use of \cite{adami2000-14} and use Shannon entropy.}

\section{Variable Length Sequences}

Physical Complexity is currently formulated for a population of sequences of the same length \cite{adami2000}, and so we will now investigate an extension to include populations of \aclp{vls}. This will require changing and re-justifying the fundamental assumptions, specifically the conditions and limits upon which Physical Complexity operates. In (\ref{complexity}) the Physical Complexity, $C$, is defined for a population of sequences of length $\ell$ \cite{adami2000}. The most important question is what does the length $\ell$ equal if the population of sequences is of variable length? The issue is what $\ell$ represents, which is the maximum possible complexity for the population \cite{adami2000}, and which we will call the \emph{complexity potential} $C_P$. The maximum complexity in (\ref{complexity}) occurs when the per-site entropies sum to zero, $\sum\limits_{i = 1}^\ell {H(i)} \tred{=} 0$, as there is no randomness in the sites (all contain information), i.e. $C \tred{=} \ell$ \cite{adami2000}. So, the \emph{complexity potential} equals the length,
\begin{equation}
C_P = \ell,
\label{comPot2}
\end{equation}
provided the population $S$ is of sufficient size for accurate calculations, as found in (\ref{popSize}), i.e. $|S|$ is equal or greater than $|D|\ell$. For a population of \aclp{vls}, $S_{V}$, the complexity potential, $C_{V_P}$, cannot be equivalent to the length $\ell$, because it does not exist. However, given the concept of minimum sample size from (\ref{popSize}), there is a length for a population of \aclp{vls}, $\ell_V$, between the minimum and maximum length, such that the number of per-site samples up to and including $\ell_V$ is sufficient for the per-site entropies to be calculated. So the \emph{complexity potential} for a population of \aclp{vls}, $C_{V_P}$, will be equivalent to its \emph{calculable} length,
\begin{equation}
\label{potential}
C_{V_P} = \ell_V.
\end{equation}

If $\ell_V$ where to be equal to the length of the longest individual(s) $\ell _{max}$ in a population of \aclp{vls} $S_{V}$, then the operational problem is that for some of the later sites, between one and $ \ell _{max}$, the sample size will be less than the population size $|S_{V}|$. So, having the length $\ell_V$ \setCap{equalling the maximum length would be incorrect}{genCap2}, as there would be an insufficient number of samples at the later sites, and therefore $\ell _V \not\equiv \ell _{max}$. So, the length for a population of \aclp{vls}, $\ell _V $, is the highest value within the range of the minimum (one) and maximum length, $1 \le \ell _V \le \ell _{\max } $, for which there are sufficient samples to calculate the entropy. A function which provides the sample size at a given site is required to specify the value of $\ell_V$ precisely,
\begin{equation}
sampleSize(i\ :site)\ :int,
\end{equation}
where the output varies between $1$ and the population size $|S_{V}|$ (inclusive). Therefore, the length of a population of \aclp{vls}, $\ell_{V}$, is the highest value within the range of one and the maximum length for which the sample size is greater than or equal to the alphabet size multiplied by the length $\ell _{V}$,
\begin{equation}
sampleSize(\ell _V ) \ge |D|\ell _V \wedge sampleSize(\ell _V + 1) < |D|\ell _V,
\label{lengthVLP}
\end{equation}
where $\ell _V$ is the length for a population of \aclp{vls}, and $\ell _{max} $ is the maximum length in a population of \aclp{vls}, $\ell _V$ varies between $ 1 \le \ell _V \le \ell _{max }$, $D$ is the alphabet and $|D| > 0$. This definition intrinsically includes a minimum size for populations of \aclp{vls}, $|D|\ell _V$, and therefore is the counterpart of (\ref{popSize}), which is the minimum population size for same length populations.

The length $\ell$ used in the limits of (\ref{persite}) no longer exists, and therefore (\ref{persite}) must be updated; so, the per-site entropy calculation for \aclp{vls} will be denoted by $H_{V}(i)$, and is,
\begin{equation}
H_V (i) = - \sum\limits_{d \in D} {p_d (i)\log _{|D|} p_d (i)},
\label{perSiteVLP}
\end{equation}
where $D$ is still the alphabet, $\ell _V$ is the length for a population of \aclp{vls}, with the site $i$ now ranging between $ 1 \le i \le \ell _V $, while the $p_d (i)$ probabilities still range between $ 0 \le p_d (i) \le 1$, and still sum to one. It remains algebraically almost identical to (\ref{persite}), but the conditions and constraints of its use will change, specifically $\ell$ is replaced by $\ell_{V}$. Naturally, $H_{V}(i)$ ranges between zero and one, as did $H(i)$ in (\ref{persite}). So, when the entropy is maximum the character found in the site $i$ is uniformly random, holding no information.

Therefore, the complexity for a population of \aclp{vls}, $C_{V}$, is the \emph{complexity potential} of the population of \aclp{vls} minus the sum, over the length of the population of \aclp{vls}, of the per-site entropies (\ref{perSiteVLP}),
\begin{equation}
\label{newComplexity}
C_{V} = \ell _V - \sum\limits_{i = 1}^{\ell _V } {H_V (i)},
\end{equation}
where $\ell_V$ is the length for the population of \aclp{vls}, and $H_V(i)$ is the entropy for a site $i$ in the population of \aclp{vls}.

\begin{figure}
    \centering
    \execute{cd images; ./pdfcrop.sh abstractAgentPopulation}
            \ifthenelse{\boolean{final}}
                {\includegraphics[width=3.5in]{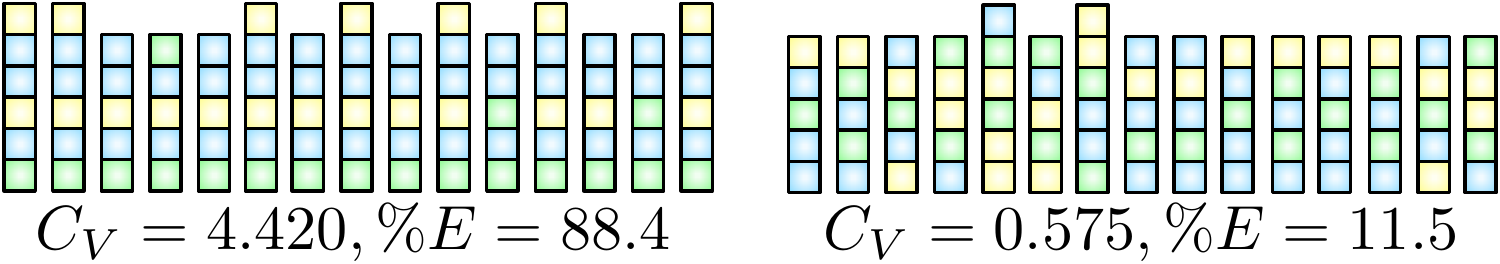}}
                {\href{file://localhost/Users/g/Desktop/PhDthesis/images/abstractAgentPopulation.graffle}{\includegraphics[width=3.33in]{images/abstractAgentPopulation-crop.pdf}}}
    \vspace{-7mm}
    \caption{\label{orgCompPop}Abstract Visualisation for Populations of Variable Length Sequences: The Physical Complexity and Efficiency values \getCap{orgCPcap}.}

    \end{figure}

Physical Complexity can now be applied to populations of \aclp{vls}, so we will consider the abstract example populations in Figure \ref{orgCompPop}. We will let a single square, \squareG{-2mm}{White}{-2.5mm}, represent a site $i$ in the sequences, with different colours to represent the different values. Therefore, a sequence of sites will be represented by a sequence of coloured squares, \linebreak \squareG{-5mm}{YellowGreenPurple}{-2.5mm}. Furthermore, the alphabet $D$ is the set \{\squaresub{-2.5mm}{Yellow}{-1mm}{-2pt}, \squaresub{-3mm}{Green}{-1mm}{-2pt}, \squaresub{-3mm}{Purple}{-1mm}{-2pt}\}, the maximum length $\ell _{max}$ is 6 and the  length for populations of \aclp{vls} $\ell _V$ is calculated as 5 from (\ref{lengthVLP}). The Physical Complexity values in Figure \ref{orgCompPop} \setCap{are consistent with the intuitive understanding one would have for the complexity of the sample populations}{orgCPcap}; the population with high Physical Complexity has a little randomness, while the population with low Physical Complexity is almost entirely random.

\section{Efficiency}

Using our extended Physical Complexity we can construct a measure showing the use of the information space, called the Efficiency $E$, which is calculated by the Physical Complexity $C_{V}$ over the complexity potential $C_{V_P }$,
\begin{equation}
E = \frac{{C_V }}{{C_{V_P } }}.
\label{efficiencyEQ}
\end{equation}
The Efficiency $E$ will range between zero and one, only reaching its maximum when the actual complexity $C_{V}$ equals the complexity potential $C_{V_{P}}$, indicating that there is no randomness in the population. In Figure \ref{orgCompPop} the populations of sequences are shown with their respective Efficiency values as percentages, and the values are as one would expect.

The complexity $C_{V}$ (\ref{newComplexity}) is an absolute measure, whereas the Efficiency $E$ (\ref{efficiencyEQ}) is a relative measure (based on the complexity $C_{V}$). So, the Efficiency $E$ can be used to compare the complexity of populations, independent of their size, their length, and whether their lengths are variable or not (as it is equally applicable to the fixed length populations of the original Physical Complexity).

\section{Clustering}
\label{cluster123}

The \emph{complexity} of a population is the \emph{clustering}, amassing of same or similar sequences, around the optimum genome \cite{begon96}. \setCap{The sequences of an evolving population will evolve, clustering around the optimal genome}{FLsingle2Cap}, assuming that its evolutionary process does not become trapped while clustering over local optima.

Clustering is indicated by the Efficiency $E$ tending to its maximum, as the population's Physical Complexity $C_{V}$ tends to the \emph{complexity potential} $C_{V_{P}}$, because an optimal sequence is becoming dominant in the population, and therefore increasing the uniformity of the sites across the population. With a global optimum, the Efficiency $E$ tends to a maximum of one, indicating that the \emph{evolving population of sequences} is tending to a \emph{set of clusters $T$ of size one},
\begin{equation}
E = \frac{C_V}{C_{V_P}} \tred{=} 1 \ as\ |T| \tred{=} 1,
\label{clusters1}
\end{equation}
assuming its evolutionary process does not become trapped at local optima. So, the \tred{Efficiency $E$} \emph{tending \tred{to its maximum}} provides a \emph{clustering coefficient}. It \emph{tends \tred{to its maximum}}, never quite reaching it, because of the mutation inherent in the evolutionary process.

If there are global optima the \setCap{Efficiency $E$ will tend to a maximum below one, because the population of sequences consists of more than one cluster, with each having an Efficiency tending to a maximum of one.}{FLmul2Cap} \setCap{The simplest scenario of clusters is \emph{pure clusters}}{FLmulCap}; \emph{pure} meaning that each cluster uses a distinct (mutually exclusive) subset of the alphabet $D$ relative to any other cluster. In this scenario the Efficiency $E$ tends to a value based on the number of clusters $|T|$, because a \emph{number} of the $p_d (i)$ probabilities at each \emph{site} in (\ref{perSiteVLP}) are the reciprocal of the number of clusters, $\frac{1}{|T|}$. So, given that the \emph{number} of the $p_d (i)$ probabilities taking the value $\frac{1}{|T|}$ is equal to the number of clusters, while the other $p_d (i)$ probabilities take a value of zero, then the per-site entropy calculation of $H_V (i)$ from (\ref{perSiteVLP}) becomes
\begin{equation}
H_V (i) = \log _{|D|} |T|,
\label{calcNumClusters}
\end{equation}
where $i$ is the site, $|D|$ is the alphabet size, and $|T|$ is the number of clusters. Hence, given (\ref{calcNumClusters}), (\ref{newComplexity}), and (\ref{potential}), then the Efficiency $E$ from (\ref{efficiencyEQ}) becomes
\begin{equation}
E \tred{=} 1 - (\log _{|D|} |T|),
\label{calcNumClusters2}
\end{equation}
where $|D|$ is the alphabet size and $|T|$ is the number of clusters. Therefore, the Efficiency $E$, the \emph{clustering coefficient}, tends to a value that can be used to determine the number of \emph{pure clusters} in an evolving population of sequences.

For a population $S$ with clusters, each cluster is a sub-population with an Efficiency $E$ tending to a maximum of one. To specify this relationship we require a function that provides the Efficiency $E$ (\ref{efficiencyEQ}) of a population or sub-population of sequences,
\begin{equation}
\mbox{\emph{efficiency(input :population) :int}}.
\end{equation}
So, for a population $S$ consisting of a set of clusters $T$, each member (cluster) $t$ is therefore a sub-population of the population $S$, and is defined as
\begin{eqnarray}
&t \in T \tred{=} & \\ \nonumber
&\left(t \subseteq S \wedge \mbox{\textit{efficiency(t)}} \tred{=} 1 \wedge |t| \approx \frac{|S|}{|T|} \wedge \sum\limits_{t \in T} {|t|} = |S|\right),&
\label{defineCluster}
\end{eqnarray}
where a cluster $t$ has an Efficiency $E$ tending to a maximum of one, and the cluster size $|t|$ is approximately equal to the population size $|S|$ divided by the number of clusters $|T|$. It is only \emph{approximately equal} because of variation from mutation, and because the population size may not divide to a whole number. These conditions are true for all members $t$ of the set of clusters $T$, and therefore the summation of the cluster sizes $|t|$ equals the size of the population $|S|$.

\tfigure{width=3.5in}{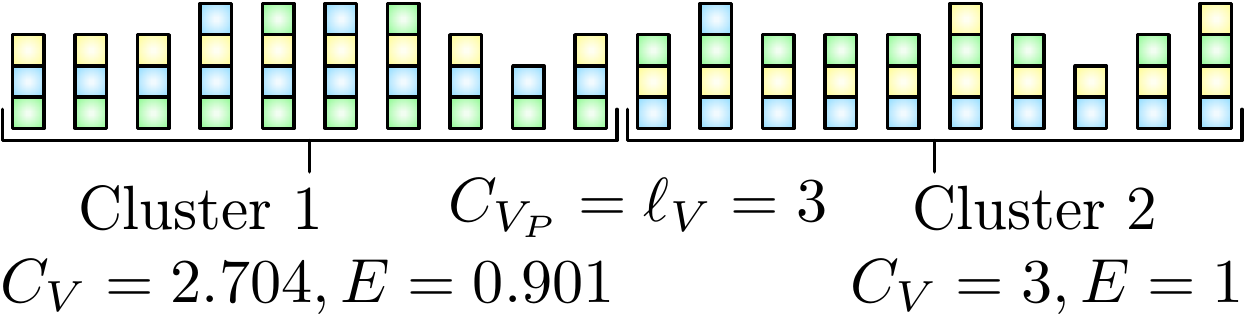}{graffle}{Population with Clusters Visible}{Visualisation for a population of sequences with global optima, which has been arranged to show the clusters present.}{-7mm}{}{}{}

A population of sequences with global optima, arranged to show the clustering, is shown in Figure \ref{populationWithClusters}. \setCap{The clusters of the population have Efficiency values tending to a maximum of one, compared to the Efficiency of the population as a whole, which is tending to a maximum significantly below one.}{popClusShowCap}

The population size $|S|$, in Figure \ref{populationWithClusters}, is double the minimum requirement specified in (\ref{lengthVLP}), so that the complexity $C_{V}$ (\ref{newComplexity}) and Efficiency $E$ (\ref{efficiencyEQ}) could be used in defining the principles of clustering without redefining the \emph{length of a population of \aclp{vls}} $\ell_{V}$ (\ref{lengthVLP}). However, when determining the variable length $\ell_{V}$ of a cluster $t$, the sample size requirement is different, specifically a cluster $t$ is a sub-population of $S$, and therefore by definition cannot have a population size equivalent to $S$ (unless the population consists of only one cluster). Therefore, to manage clusters requires a reformulation of $\ell _V$ (\ref{lengthVLP}) to \tred{the highest value within the range of one and the maximum length for which the sample size is greater than or equal to the alphabet size multiplied by the length and divided by the number of clusters,}
\begin{equation}
\ell _V = \left(
\begin{array}{l}
sampleSize(\ell _V ) \approx \frac{|D|\ell _V }{|T|} \wedge \\
\vspace{-3mm}\white{.} \\
sampleSize(\ell_{V} + 1) < \frac{|D|\ell _V }{|T|} \\
\end{array}\right),
\label{clustersSampleSize}
\end{equation}
where $\ell _{max} $ is the maximum length in a population of \aclp{vls}, $\ell _V$ varies between $ 1 \le \ell _V \le \ell _{max }$, $D$ is the alphabet, $|D| > 0$, and $T$ is the set of clusters in the population $S$.

A population with clusters will always have an Efficiency $E$ tending towards a maximum significantly below one. Therefore, managing populations with clusters requires a reformulation of the Efficiency (\ref{efficiencyEQ}) to
\begin{equation}
E_{c} (S) = \left\{
\begin{array}{cl}
\frac{C_V }{C_{V_P}} & \mbox{if $|T| = 1$} \\
\vspace{-2mm}\white{.} & \white{.}\\
\frac{\sum\limits_{t \in T} {E_{c} (t)}}{|T|} & \mbox{if $|T| > 1$}
\end{array}\right.,
\label{efficiencyMultiple}
\end{equation}
where $t$ is a cluster, and a member of the set of clusters $T$ of the population $S$. So, the Efficiency $E_{c}$ is equivalent to the Efficiency $E$ if the population consists of only one cluster, but if there are clusters then the Efficiency $E_{c}$ is the average of the Efficiency $E$ values of the clusters.

\section{Simulation and Results}

A simulated population $S$ of sequences, $[s_1, s_1, s_2, ...]$, was evolved to a \emph{selection pressure} $R$. A dynamic population size was used to ensure exploration of the available search space, which increased with the average length of the population. The optimal sequences were evolved to a \emph{fitness function} generated from the \emph{selection pressure} $R$. Each sequence of the population consisted of a list of sites, $[i_1, i_2, ...]$, while the \emph{selection pressure} consisted of a list of attributes, $[r_1, r_2, ...]$. So, the \emph{fitness function} for evaluating a sequence $s$, relative to the \emph{selection pressure} $R$, was
\begin{equation}
fitness(s,R) = \frac{1}{1 + \sum_{r \in R}{|r-i|}},
\label{ff}
\end{equation}
where $i$ is a site of a sequence $s$ measured against corresponding site $r$ of the \emph{selection pressure} $R$. Equation \ref{ff} was used to assign \emph{fitness} values between 0.0 and 1.0 to each individual of the current generation of the population, directly affecting their ability to replicate into the next generation. The evolutionary computing process was encoded with a low mutation rate, a fixed selection pressure and a non-trapping fitness function (i.e. did not get trapped at local optima). The type of selection used was \emph{fitness-proportional} and \emph{non-elitist}, \emph{fitness-proportional} meant that the \emph{fitter} the individual the higher its probability of surviving to the next generation. \emph{Non-elitist} meant that the best individual from one generation was not guaranteed to survive to the next generation; it had a high probability of surviving into the next generation, but it was not guaranteed as it might have been mutated. \emph{Crossover} (recombination) was then applied to a randomly chosen 10\% of the surviving population, a \emph{one-point crossover}, by aligning two parent individuals and picking a random point along their length, and at that point exchanging their tails to create two offspring. \emph{Mutations} were then applied to a randomly chosen 10\% of the surviving population; \emph{point mutations} were randomly located, consisting of \emph{insertions}, \emph{replacements}, and \emph{deletions}. The issue of bloat was controlled by augmenting the \emph{fitness function} with a \emph{parsimony pressure} which biased the search to shorter sequences, evaluating longer than average sequences with a reduced \emph{fitness}, and thereby providing a dynamic control limit which adapted to the average length of the ever-changing evolving populations.

\vspace{5mm}

Figure \ref{phycom} shows, for a typical evolving population, the Physical Complexity $C_V$ (\ref{newComplexity}) for \aclp{vls} and the \emph{maximum fitness} $F_{max}$ over the generations. It shows that the fitness and our extended Physical Complexity; both increase over the generations, synchronised with one another, until generation 160 when the \emph{maximum fitness} tapers off more slowly than the Physical Complexity. At this point the optimal length for the sequences is reached within the simulation, and so the advent of new fitter sequences (of the same of similar length) creates only minor fluctuations in the Physical Complexity, while having a more significant effect on the \emph{maximum fitness}. The similarity of the graph in Figure \ref{phycom} to the graphs in \cite{adami20002} confirms that the Physical Complexity measure has been successfully extended to \aclp{vls}.

\tfigure{width=3.5in}{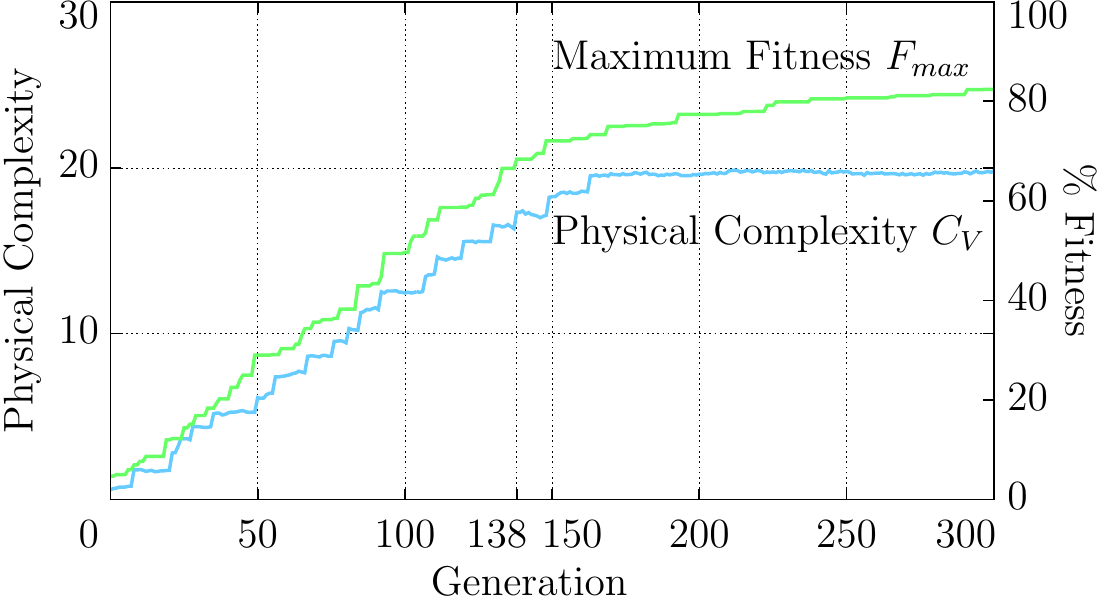}{graph}{Graph of Physical Complexity and Maximum Fitness over the Generations}{The Physical Complexity for \aclp{vls} increases over the generations, showing short-term decreases as expected, such as at generation 138.}{-7mm}{}{}{}

\subsection{Efficiency}

Figure \ref{newphycomvis} is a visualisation of the simulation, showing two alternate populations that were run for a thousand generations, with the one on the left from Figure \ref{phycom} run under normal conditions, while the one on the right was run with a non-discriminating selection pressure \cite{kimura:ntm}. Each multi-coloured line represents a sequence, while each colour represents a site. \setCap{The visualisation shows that our Efficiency $E$ accurately measures the complexity of the two populations.}{largeVisCap}

\tfigure{width=3.5in}{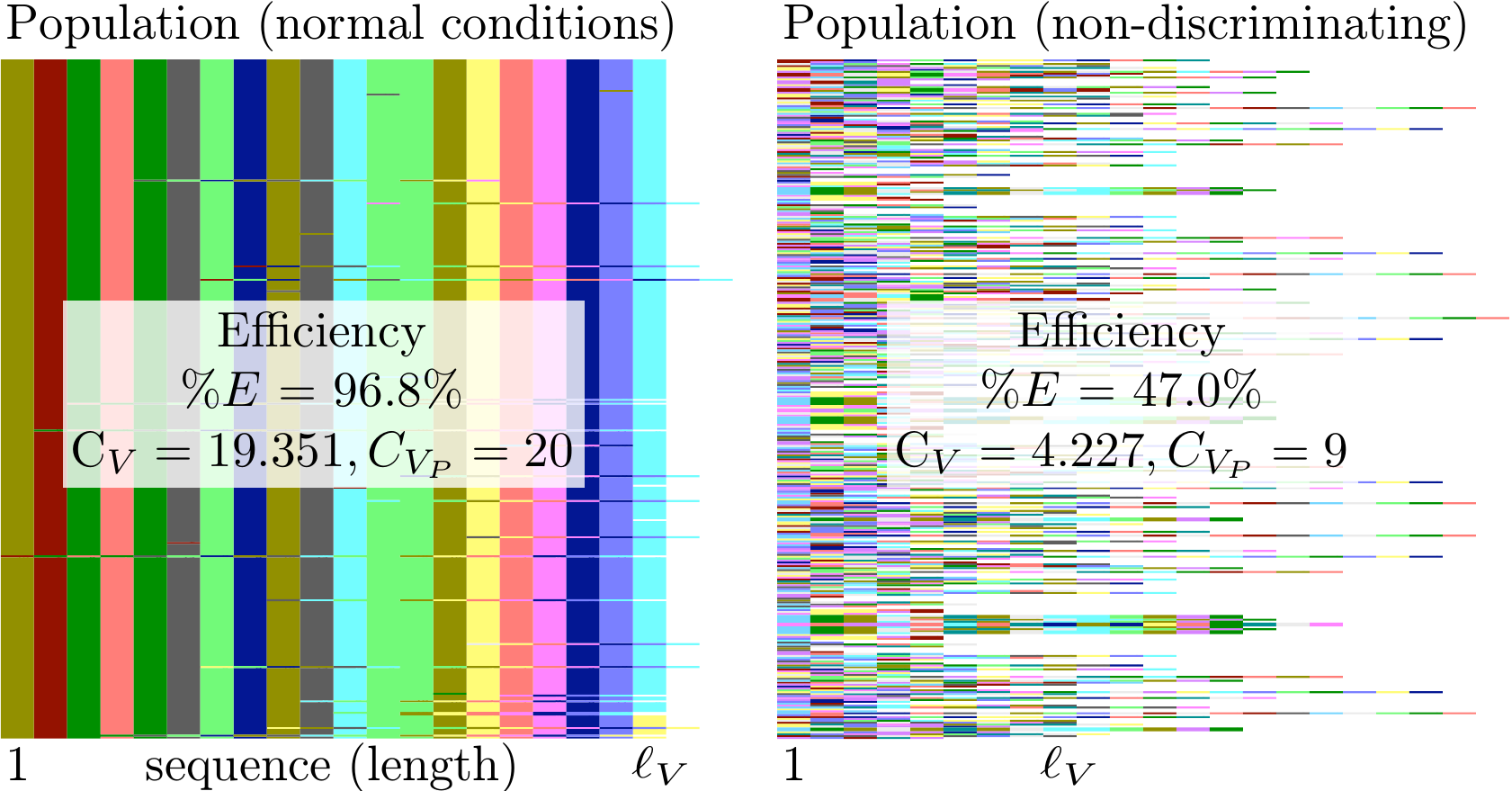}{graffle}{Visualisation of Evolving Populations at the 1000th Generation}{The population on the left from Figure \ref{phycom} was run under normal conditions, while the one on the right was run with a non-discriminating selection pressure.}{-7mm}{!t}{}{}

Figure \ref{efficiency} shows the Efficiency $E$ (\ref{efficiencyEQ}), over the generations, for the population from Figure \ref{phycom}. \setCap{The Efficiency tends to a maximum of one, indicating that the population consists of one cluster}{graph32cap}, which is confirmed by the visualisation of the population in Figure \ref{newphycomvis} (left).

\tfigure{width=3.5in}{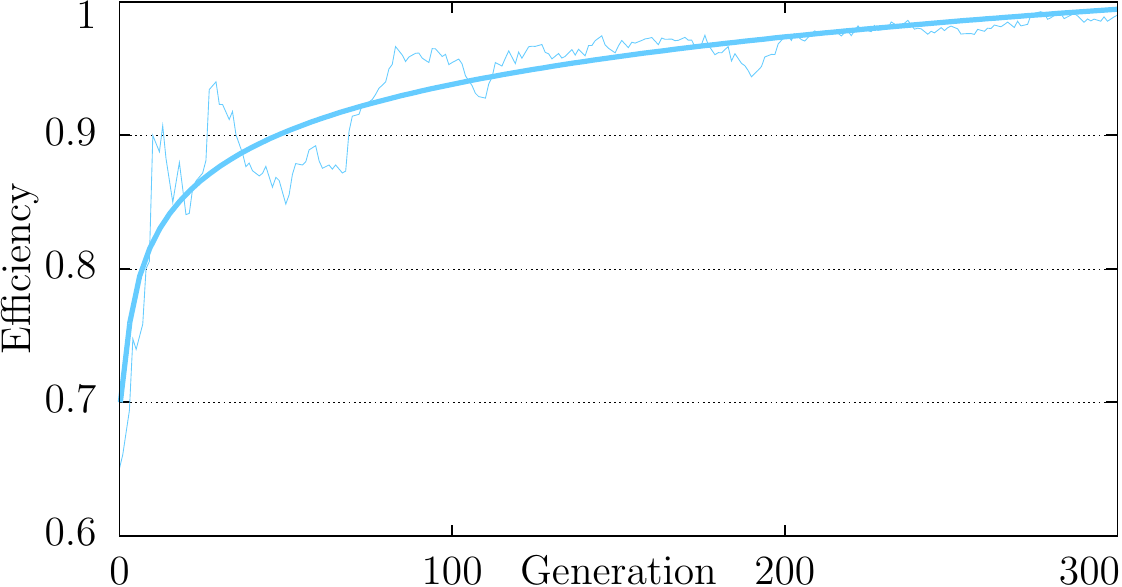}{graph}{Graph of Population Efficiency over the Generations for the population from Figure \ref{phycom}}{\getCap{graph32cap}.}{-7mm}{!t}{}{}

\subsection{Clustering}

\tfigure{width=3.5in}{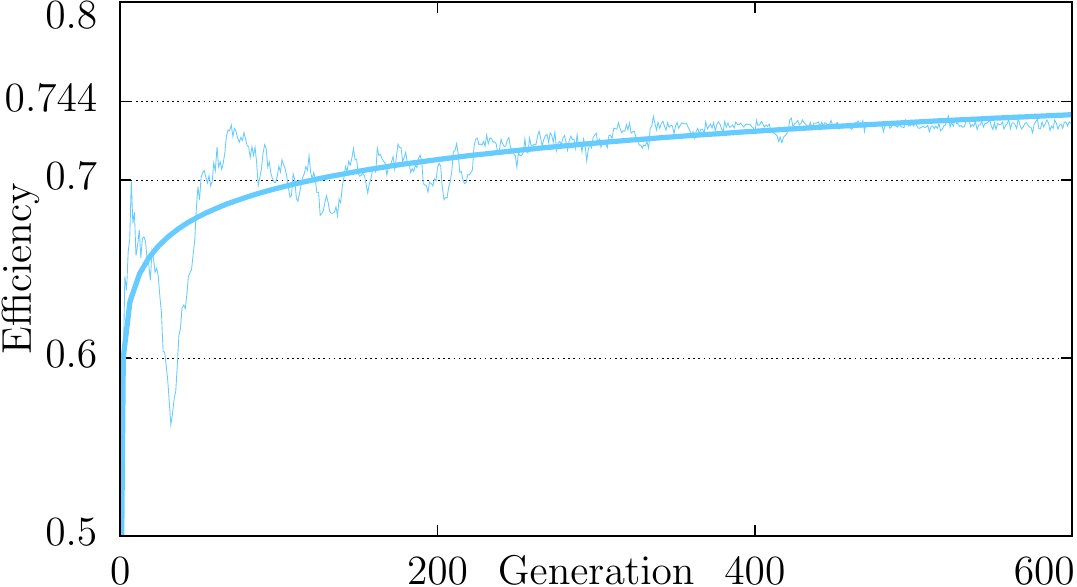}{graph}{Graph of the Clustering Coefficient over the Generations}{The Efficiency oscillated to 0.744, as expected from (\ref{calcNumClusters2}) given the alphabet size was fifteen, $|D|$=15, and the number of clusters was two, $|T|$=2, indicating more than one cluster.}{-7mm}{!h}{}{}

To further investigate the complexity of evolving populations, we simulated a typical population with a multi-objective \emph{selection pressure} that had two independent global optima, and so the potential to support two \emph{pure clusters} (each cluster using a unique subset of the alphabet $D$). The graph in Figure \ref{coefficient} shows the Efficiency $E$ over the generations acting as a \emph{clustering coefficient}, oscillating \setCap{around the included best fit curve, quite significantly at the start, and then decreasing as the generations progressed.}{graph4cap} The Efficiency tended to 0.744, as expected from (\ref{calcNumClusters2}) given the alphabet size was fifteen, $|D|$=15, and the number of clusters was two, $|T|$=2. \tred{It indicated the occurrence of} clustering, while the value it tended to indicated, as expected, the presence of two clusters in the population. A visualisation of the population is shown in Figure \ref{newCluster}, in which the \setCap{sequences were grouped to show the two clusters}{visClustersCap}. As \setCap{expected from (\ref{defineCluster}) each cluster had a much higher Physical Complexity and Efficiency compared to the population as a whole. However, the Efficiency $E_{c}$}{visClusters2Cap} is immune to the clusters and therefore \setCap{calculated the}{visClusters3Cap} complexity of the population correctly.

\tfigure{width=3.5in}{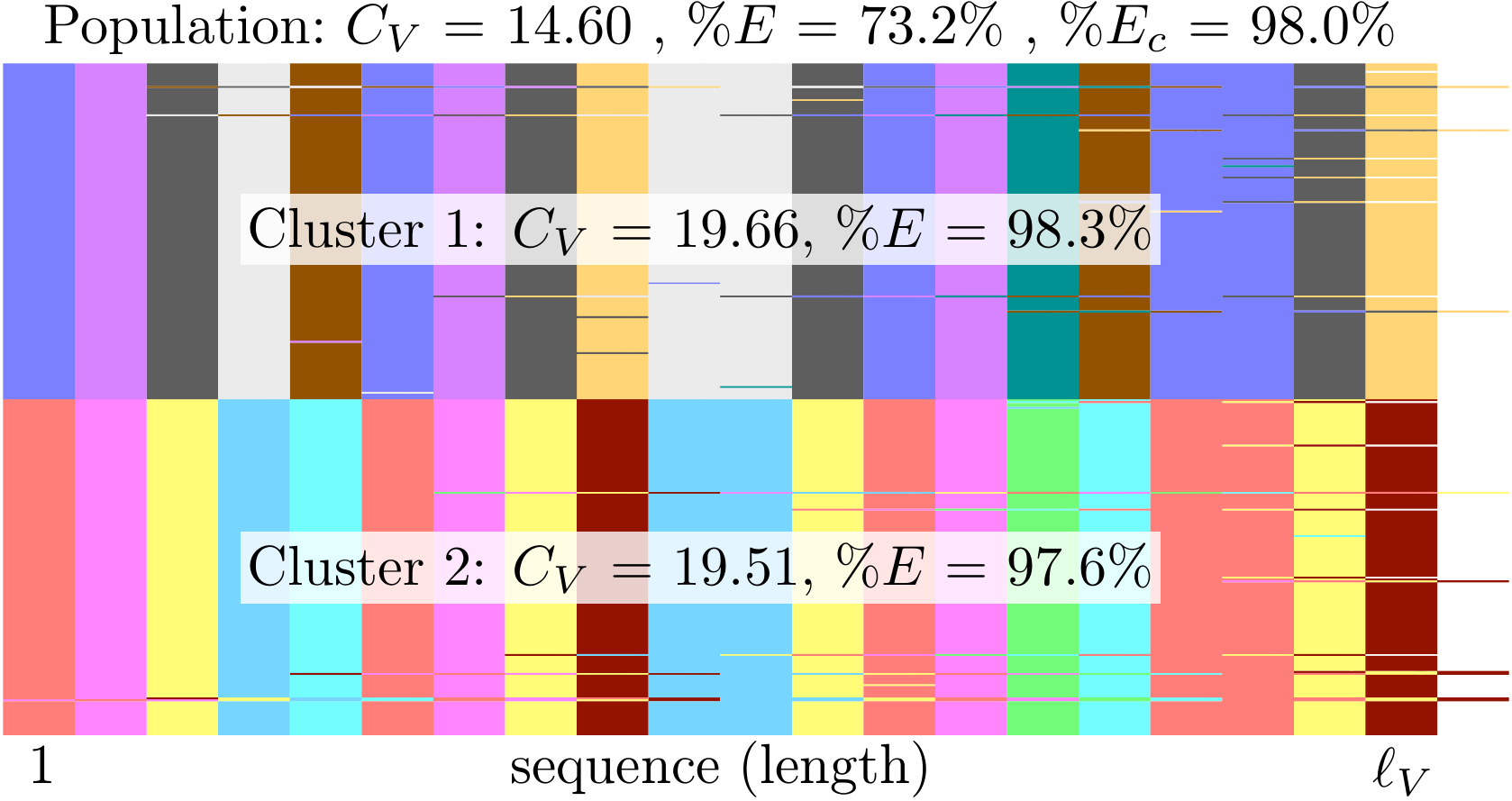}{graffle}{Visualisation of Clusters in an Evolving Population at the 1000th Generation}{The \getCap{visClustersCap}.}{-7mm}{}{}{}

\section{Conclusions}

Physical Complexity \cite{adami20002} is based upon the individuals of a population within the context of their environment, but was only formulated for populations of fixed length. However, this was not a fundamental property of its definition \cite{adami20002}, and so we chose to extend it to include populations of variable length sequences. We then built upon this to construct a variant of the Physical Complexity called the Efficiency, because it was based on the efficiency of information storage, which was then used to develop an understanding of clustering and atomicity in evolving populations with multi-objective \emph{selection pressures}. The \emph{clustering coefficient} defined by the Efficiency tend\tred{ing to its maximum,} not only indicates clustering, but can also determine the number of clusters (for \emph{pure clusters}).

We then investigated our extended Physical Complexity through experimental simulations, with the results consistent with the original, confirming a successful algebraic reformulation to include populations of variable length sequences. We then investigated the Efficiency, which performed as expected, confirmed by the numerical results and the visualisations matching our intuitive understanding. We then applied the Efficiency to the determination of clusters for populations with multi-objective \emph{selection pressures}. The numerical results, combined with the visualisations of multi-cluster populations, confirmed the ability of the Efficiency to act as a \emph{clustering coefficient}, not only indicating the occurrence of clustering, but also the number of clusters. Finally, we confirmed that the \emph{Efficiency $E_{c}$} (for populations with clusters) could calculate correctly the Physical Complexity for variable length populations with clusters.

Overall, Physical Complexity has been extended to include populations of variable length, with our Efficiency definition providing a macroscopic value to characterise the level of Physical Complexity. Furthermore, our Efficiency $E_{c}$ provides a normalised \emph{universally applicable} macroscopic value to characterise the complexity of any population, independent of clustering, atomicity, length (variable or same), and size. Most importantly, the understanding and techniques we have developed have applicability as wide as the original Physical Complexity, which has been applied from \acs{DNA} \cite{adami2000} to simulations of self-replicating programmes \cite{adami20032}.

\section{Acknowledgments}

The authors would like to thank the following for encouragement and suggestions; Dr Paolo Dini of the London School of Economics and Political Science, and Dr Christoph Adami of Keck Graduate Institute.

\nocite{basharin,kolmogorov,adami2000-14,thomasjr1971goc,vonbertalanffy1973gst,kimura:ntm,zurek1990aic,li1997ikc,adami1998ial,adami2000,adami20002,gregory2001cco,adami2003,jaynes2003ptl,adami20032}

\bibliographystyle{abbrv}
\bibliography{references,myRefs}
\end{document}